\documentclass[url,amsmath,epsfig,graphicx,12pt,prl,superscriptaddress,preprint]{revtex4}

\usepackage{graphicx}
\usepackage{dcolumn}
\usepackage{bm}
\newcommand{\rv}{{\bf r}}
\newcommand{\xig}{{\xi_g}}
\newcommand{\sigmaeff}{{\sigma_{\rm eff}}}
\newcommand{\etal}{{\em et al.}}

\begin{document}

\title{Non-equilibrium sedimentation of colloids on the particle scale}

\author{C. Patrick Royall}
\affiliation{Institute of Industrial Science, 
  University of Tokyo, 4-6-1 Komaba, Meguro-ku, Tokyo 153-8505, Japan}
\affiliation{Soft Condensed Matter, Debye Institute, Utrecht University, 
  Princetonplein 5, 3584CC, Utrecht, The Netherlands}

\author{Joachim Dzubiella}
\affiliation{Physics Department (T37), Technical University Munich,
  James Franck Stra\ss e, D-85748 Garching, Germany}

\author{Matthias Schmidt}
\affiliation{H.H. Wills Physics Laboratory, University of Bristol,
  Tyndall Avenue, Bristol BS8 1TL, UK}

\affiliation{Institut f{\"u}r Theoretische Physik II,
  Heinrich-Heine-Universit{\"a}t D{\"u}sseldorf,
  Universit{\"a}tsstra{\ss}e 1, D-40225 D{\"u}sseldorf, Germany}

\author{Alfons van Blaaderen}
\affiliation{Soft Condensed Matter, Debye Institute, Utrecht University, 
  Princetonplein 5, 3584CC, Utrecht, The Netherlands}


\date{24 March 2007}



\begin{abstract}
   We investigate sedimentation of model hard sphere-like colloidal
   dispersions confined in horizontal capillaries using laser scanning
   confocal microscopy, dynamical density functional theory, and
   Brownian dynamics computer simulations. For homogenized initial
   states we obtain quantitative agreement of the results from the
   respective approaches for the time evolution of the one-body
   density distribution and the osmotic pressure on the walls.  We
   demonstrate that single particle information can be obtained
   experimentally in systems that were initialized further
   out-of-equilibrium such that complex lateral patterns form.
\end{abstract}

\maketitle

The process of dispersed particles settling under gravity is a
prevalent phenomenon in colloidal dispersions, relevant for isolation
of cells from blood \cite{boyum1968}, separation of macromolecules
\cite{wang2002}, industrial filtration \cite{tiller1995}, and the
shelf-life of paint \cite{dolez2002}.  As demonstrated by the
beautiful pioneering work of Perrin \cite{perrin}, particles with
buoyant mass $m$ that are exposed to a gravitational field with
acceleration~$g$ can be viewed as being subject to a (linearly)
changing local chemical potential $\mu(z) = -mgz +{\rm const}$, where
$z$ is the height variable. This provides a means to access isotherms
and e.g.\ study equilibrium phase behavior, provided that the
gravitational length, $\xig=k_BT/(m g)$, where $k_B$ is the Boltzmann
constant, and $T$ is temperature, is large compared to the
inter-particle correlation length, but small compared to the vertical
container size $L$ \cite{piazza1993,rutgers1996}. In equilibrium a
pressure gradient that originates from an inhomogeneous density
profile, $\rho(z)$, balances gravity and for a dilute dispersion ideal
gas-like behavior results, $\rho(z) \propto \exp(-z/\xig)$.  In
concentrated systems pronounced density oscillations, indicative of
liquid layering, occur near the lower wall~\cite{biben1993}.

Understanding the {\em time evolution} of a given initial state
towards sedimentation-diffusion equilibrium is a difficult problem due
to the interplay of direct and (solvent-mediated) hydrodynamic
interactions between the particles \cite{janDhont}.  An isolated
sphere of radius $a$ in a solvent of dynamical viscosity $\eta$
sediments at an average (Stokes) velocity $u_0 = \Gamma_0 mg$, where
the mobility is $\Gamma_0=(6\pi\eta a)^{-1}$ for no-slip boundary
conditions on the sphere.  Two timescales are of relevance, that of
diffusion is the Brownian time it takes a particle to diffuse over a
distance of one radius, $\tau_B=3\pi\eta\sigma^3/(4k_BT)$, and that of
gravity is the time it takes a particle to sediment over one radius,
$\tau_S=\sigma/(2u_0)$, where $\sigma=2a$ is the particle
diameter. The ratio between both is the Peclet number, $\text{Pe} =
\tau_B/\tau_S = a/\xig$, that characterizes the type of dynamics.
Previous experiments were carried out using sedimenting granular
matter \cite{segre2001} where the effects of diffusion are small
($\text{Pe}\gg 1$), and using relatively small colloids
\cite{janDhont,albertPhilipse} where diffusion dominates
($\text{Pe}\ll 1$).  Steady-state in the intermediate regime was
recently investigated with computer simulations \cite{ardLouis2004}
showing that hydrodynamic backflow effects are relevant but depend
little on Pe and that the velocity correlations are similar to those
near equilibrium.

In this Letter we investigate full non-equilibrium sedimentation of
dispersions inside horizontal planar slits with plate distance $L$ and
$\text{Pe}\sim 1$ using experiment, theory and simulation, all of
which are capable of resolving the microscopic structure of the
dispersion on the particle level. We obtain quantitative agreement of
results from the respective approaches for the strongly inhomogeneous
density profiles. This contributes to our understanding of the effects
of particle correlations and confinement on the non-equilibrium time
evolution of sedimentation and could be relevant for transport of
cells in microfluidic devices \cite{yao2004}, bacteria in groundwater
\cite{harvey1997} and in porous media~\cite{wan1995}.

\begin{center} 
  \begin{figure}[t]
    \includegraphics[width=0.30\columnwidth]{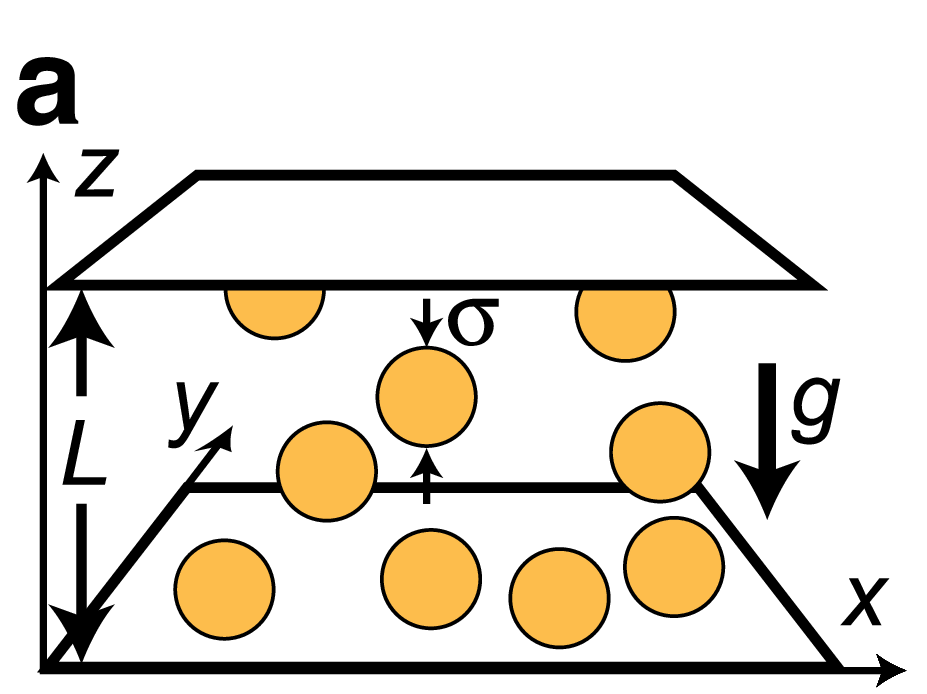}
    \includegraphics[width=0.68\columnwidth]{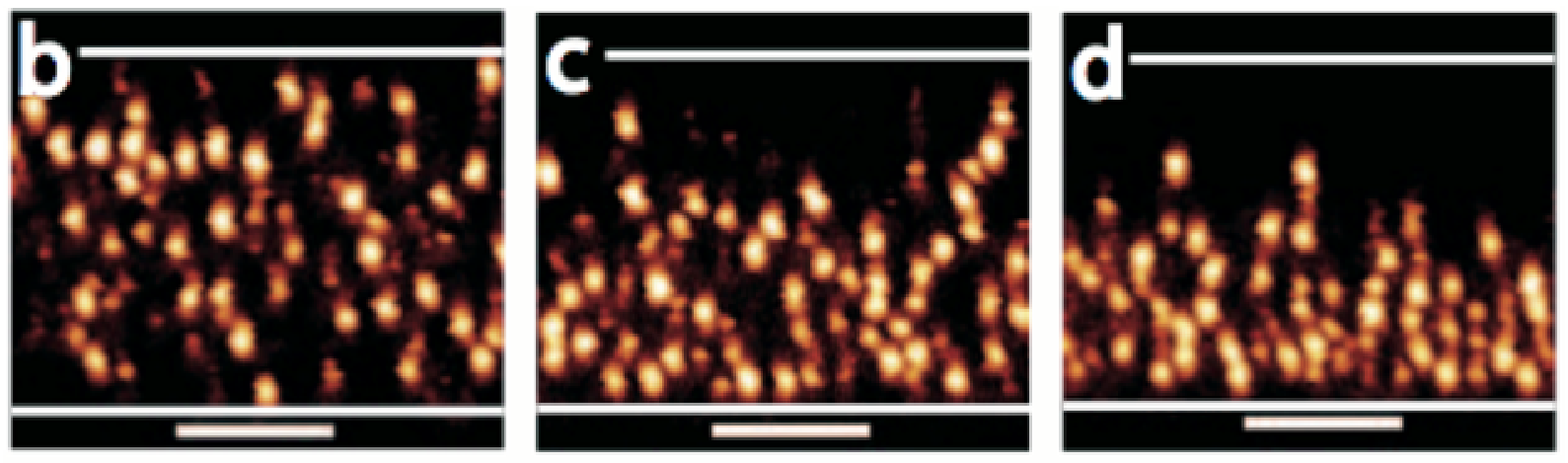}
    \caption{Non-equilibrium sedimentation of colloids.  (a)
      Illustration of colloidal spheres of diameter $\sigma$ under the
      influence of gravity $g$ and vertically confined between two
      walls with separation distance $L$. The vertical coordinate is
      $z$; the horizontal coordinates are $x$ and $y$.  (b), (c), (d),
      Time series of confocal micrographs of system A (see Tab.\
      \ref{table1} for parameters) taken in the (vertical) $xz$ plane
      at times $t/\tau_B=3, 26, 200$, respectively. The scale bars
      denote 20 $\mu$m; the horizontal lines indicate the position of
      the walls.}
    \label{figPix}
  \end{figure}
\end{center}

Our experimental system \cite{paddy2005} consists of micron-sized,
sterically stabilised, fluorescently labelled Poly methyl methacrylate
particles \cite{bosma2002} (PMMA), dispersed in a mixture of
cyclohexyl bromide (CHB) and cis-decalin with 260 $\mu$M tetrabutyl
ammonium bromide (TBAB) salt to screen the (weak) electrostatic
interactions between the colloids, resulting in a Debye screening
length of $\sim$100~nm \cite{paddy2005}; see Tab.\ \ref{table1} for a
summary of all relevant parameters. We tailored the density difference
between solvent and particles to control the buoyant mass~$m$ and
hence Pe.  Initially dry capillaries were filled with the suspension;
this is a very rapid process due to large capillary forces between the
walls and the solvent.  The capillaries were laid flat such that the
colloids sediment towards the bottom wall, and imaged with a confocal
laser scanning microscope, see Fig.\ \ref{figPix}a for an
illustration. Confocal scans were made in three dimensions and time
series are presented as slices in either vertical ($xz$, see Fig.\
\ref{figPix}b-d) or horizontal ($xy$) planes. Coordinates of the
particles were tracked in three dimensions \cite{royall2003}, yielding
the colloid density profile, $\rho(z,t)$, as a function of the
altitude $z$ and time $t$.  We present results for $L/\sigma = 12$
(system A) and $L/\sigma = 38.5$ (system B).

\begin{center} 
  \begin{figure}[htb]
   \vspace{-12mm}
    \includegraphics[width=0.6\columnwidth]{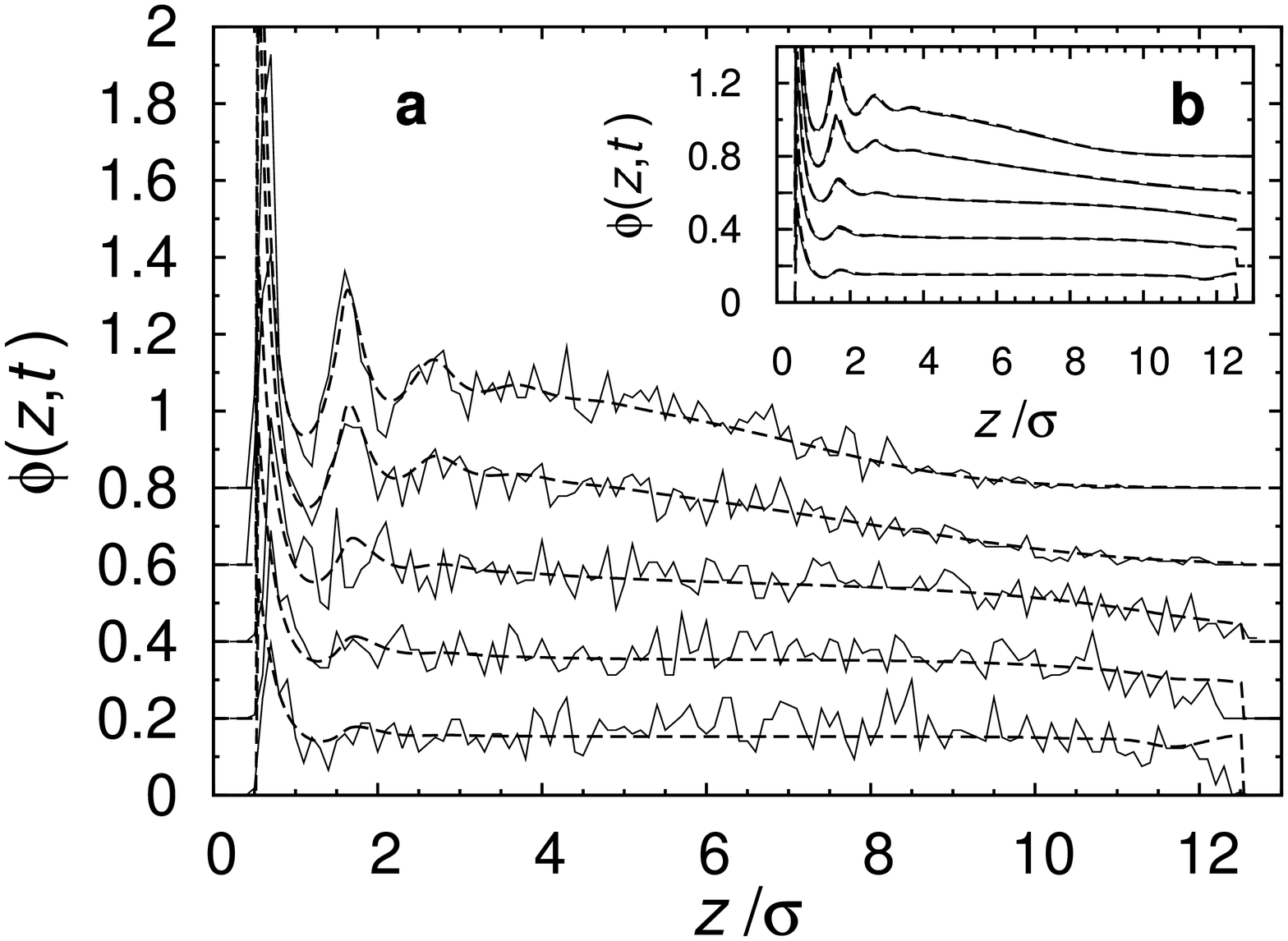}
    \includegraphics[width=0.62\columnwidth]{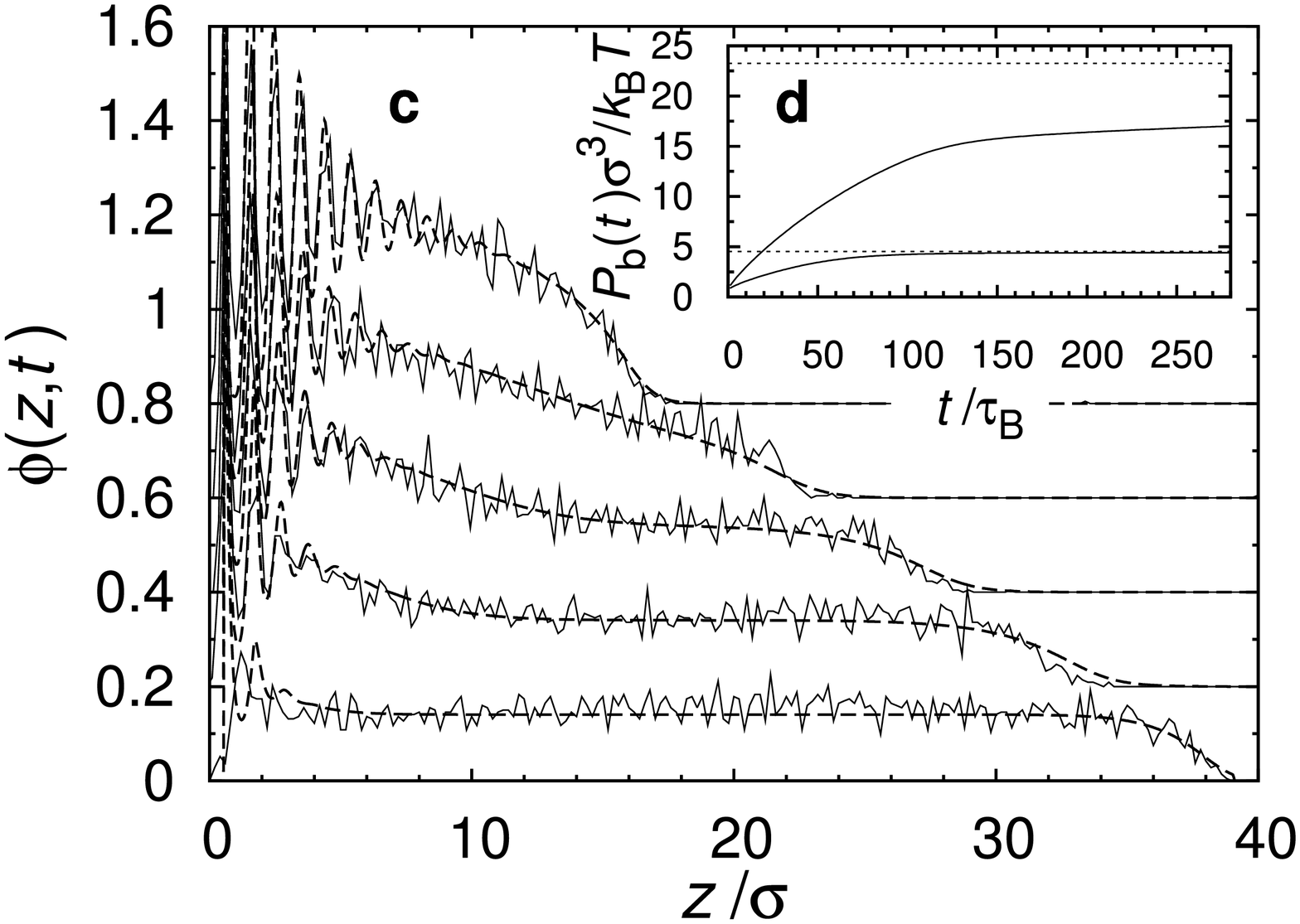}
    \caption{Time evolution of the density distribution. (a) The local
      packing fraction $\phi(z,t)=\rho(z,t)\pi \sigma^3/6$ as a
      function of the (scaled) height coordinate $z/\sigma$ for
      (scaled) times $t/\tau_B=3, 7, 14, 39, 100$ (from bottom to top)
      for system A. Shown are results from experiment (full lines) and
      from DDFT using a density-dependent mobility (dashed lines). The
      curves are vertically offset by 0.2 for clarity. (b) Same as (a)
      but obtained from BD computer simulations (full lines) and DDFT
      (dashed lines); in both approaches a constant mobility is
      used. (c) Same as (a) but for system B for times $t/\tau_B=7,
      26, 44, 63, 255$ (from bottom to top).  Time $t=0$ is set such
      that the results for the earliest time match; the theoretical
      curves are scaled with a value of $\tau_B$ that is 20\% larger
      than the experimental value. The experimental result for $t=255
      \tau_B$ is compared to the theoretical equilibrium profile.  (d)
      The osmotic pressure that the particles exert on the lower wall,
      $P_{\rm b}(t)$, as a function of the scaled time, $t/\tau_B$, as
      obtained from DDFT for system A (lower solid line) and system B
      (upper solid line).  The asymptotic values for $t\to\infty$ are
      indicated by dotted lines. Results are obtained using a
      density-dependent mobility.  }
    \label{figProfiles}
  \end{figure}
\end{center}

Experimental results for the time evolution of the sedimentation
profile for system A are shown in Fig.\ \ref{figProfiles}a. The inital
density distribution is almost flat, save for a small peak at the
lower wall. This peak grows in time and two more layers appear. The
final profile exhibits clear second and third layers and decays
smoothly for higher altitudes.  We have observed no pronounced changes
in the lateral ($xy$) structure, nor strong fluctuations in particle
velocity, in accord with the findings of Ref.\
\onlinecite{ardLouis2004} and contrasting with the case of strongly
driven (steady-state) sedimentation \cite{segre2001}.


A recent and very promising theory for the time evolution of the
one-body density, $\rho(\rv,t)$, where $\rv$ is the position
coordinate, is dynamical density functional theory
\cite{marini1999,archer2004} (DDFT):
\begin{align}
  \frac{\partial \rho}{\partial t} &= \nabla \cdot \left(
  \Gamma \rho \nabla \frac{\delta F[\rho]}{\delta \rho} \right),
  \label{EQddft}
\end{align}
where $\Gamma$ is the mobility, $F[\rho]$ is taken to be the
equilibrium Helmholtz free energy functional, $F[\rho]= k_BT\int d\rv
\rho(\rv)[\ln(\rho(\rv,t)\Lambda^3)-1] + F_{\rm exc}[\rho] + \int d\rv
\rho(\rv) V_{\rm ext}(\rv) $, where the first term is the Helmholtz
free energy of the ideal gas, $\Lambda$ is the (irrelevant) thermal
wavelength, $F_{\rm exc}[\rho]$ is the excess (over ideal gas)
contribution to the free energy, and the external potential in the
present case is $V_{\rm ext}(\rv)=k_BT z/\xi_g$ for
$\sigma/2<z<L-\sigma/2$ and $\infty$ otherwise.  The interparticle
interactions are described by the excess free energy functional
$F_{\rm exc}[\rho]$, which we take to be that of hard spheres of
diameter $\sigmaeff>\sigma$ to model the (strongly screened)
electrostatic repulsion. We use Rosenfeld's successful
fundamental-measures theory \cite{rosenfeld1989} to approximate
$F_{\rm exc}[\rho]$.  As a simple treatment of hydrodynamic
interactions we take into account the reduction of $\Gamma$ upon
increasing density with the result by Hayakawa and Ichiki
\cite{hayakawa1995}, $\Gamma_{\rm HI}(\phi)/\Gamma_0=
(1-\phi)^3/[1+2\phi+1.492\phi(1-\phi)^3]$, where $\phi=4\pi \rho
a^3/3$ is the colloid packing fraction, and we generalize to strongly
inhomogeneous situations by using the local packing fraction
$\bar\phi(\rv,t)$, obtained as the convolution of the bare density
profile, $\rho(\rv,t)$, with a weight function characteristic of the
particle volume \cite{rosenfeld1989}. We determine values for $\xig$
and $\sigmaeff$ such that the theoretical result for the equilibrium
sedimentation profile [obtained from solving $\delta F[\rho]/\delta
\rho(\rv)=\mu$ for $\rho(\rv)$, where $\mu$ is the chemical potential]
matches its experimental counterpart.  The resulting agreement is very
good indeed, see the top pair of curves in Fig.\ \ref{figProfiles}a.
In addition we have checked that values for $\sigmaeff$ are consistent
with our experimental results for the radial distribution function
(not shown) and that the values for $\xig$ are compatible with those
obtained from estimating $m$ from sample composition. We assume
$\rho(\rv,t)=\rho(z,t)$, take $\rho(z,t=0)$ to be the equilibrium
profile for hard spheres between hard walls without gravity, and
integrate Eq.\ (\ref{EQddft}) forward in time.  The results for
$\rho(z,t)$ compare very well with the experimental data, including
the time evolution of the layering at the bottom which was not
captured in earlier treatments \cite{janDhont}, where monotonic
density profiles were obtained -- as is appropriate on a macroscopic
scale.


We further test the DDFT by comparing results for $\rho(z,t)$ to those
from BD simulations. To correct for the finite resolution of the
simulations, we have used an intrinsic clock to \emph{measure} time:
The clock ticks only when a trial move is accepted, but not when a
trial move is rejected due to hard core overlap. We have extensively
tested this approach against time step extrapolation
\cite{cichocki1990} and are confident that it is accurate using time
steps of $10^{-3}\tau_B$, hence offering significant advantages in
computational efficiency.  We use a constant mobility $\Gamma_{\rm
HI}(\phi_0)$ both in BD and DDFT, with $\phi_0$ being the overall
packing fraction. The agreement between results from theory and
simulation is remarkable, see Fig.\ \ref{figProfiles}b, establishing
for the first time that DDFT is accurate for describing Brownian
dynamics of hard spheres in three dimensions; this is both of
fundamental and practical importance due to the high computational
efficiency of the scheme.

System B serves to investigate the crossover from strongly confined to
macroscopic systems. In the early stages we find a large region at
intermediate heights where the density remains nearly constant,
reminiscent of batch settling \cite{russel}, as well as a pronounced
upper interface of the sediment towards an essentially particle-free
supernatant, see Fig.\ \ref{figProfiles}c. For long times pronounced
layering at the bottom of the container develops; we have analyzed the
lateral structure in the experiment and have ascertained that these
layers remain liquid-like, with no signs of crystallization, which is
expected to occur for yet more layers \cite{hoogenboom2003}.  We
achieve very good agreement of the experimental and theoretical
results for $\rho(z,t)$, albeit with a small mismatch in time scales:
The theoretical results are shown for a value of $\tau_B$ that is 20\%
larger than the experimental value $\tau_B = 29.3$ s, obtained from
measurements of particle size and viscosity. The discrepancy might be
due to particle size polydispersity ($\sim 5\%$) becoming more
important in the case of the larger value of Pe in system B, and not
taken into account in the DDFT.

The detailed structural information contained in $\rho(z,t)$ enables
us to investigate the time evolution of important macroscopic
quantities such as the osmotic pressure \cite{biben1993} exerted by
the particles on the bottom and top walls, $P_{\rm b}(t) = k_BT
\rho(z=\sigma/2,t)$ and $P_{\rm t}(t) = k_BT \rho(z=L-\sigma/2,t)$,
respectively. Results shown in Fig.\ \ref{figProfiles}d indicate a
sharp initial increase in $P_{\rm b}(t)$ with a crossover to slow
relaxation for long times towards its equilibrium value, whereas
$P_{\rm t}(t)$ very rapidly decays (not shown). Note that in
equilibrium $P_{\rm b}-P_{\rm t} = k_BT \nu/\xig$, where $\nu$ is the
number density per area.

\begin{center} 
  \begin{figure}[htb]
    \includegraphics[width=0.5\columnwidth]{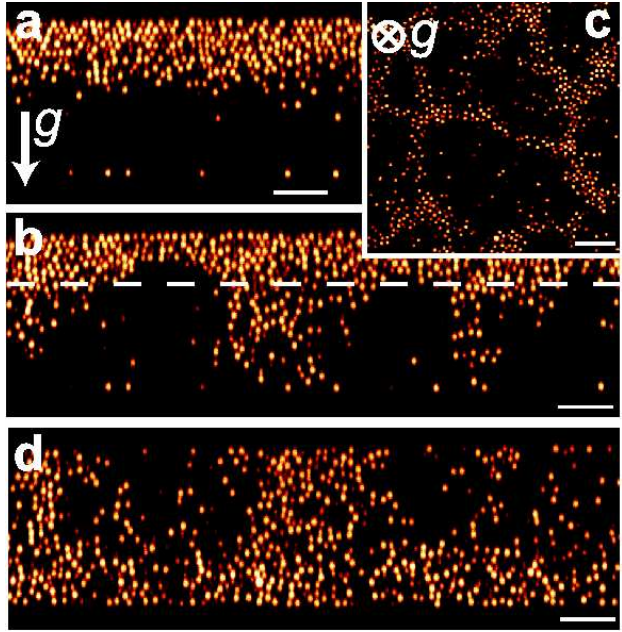}
    \caption{Rayleigh-Taylor-like instability.  An equilibrated sample
      is turned upside down and left to sediment ((a), vertical $xz$
      plane at time $t=2 \tau_B$). A strong fingering-like
      inhomogeneity develops quickly ((b), vertical $xz$ plane at
      $t=5\tau_B$), accompanied by maze-like lateral pattern formation
      ((c) horizonal $xy$ plane at time $t=9 \tau_B$).  Subsequently
      particle diffusion equilibrates the system ((d) vertical
      $xz$-plane at time $t=11\tau_B$), similar to the time evolution
      shown in Fig.~\ref{figPix}c-d.  The dashed line in (b) indicates
      the height of the $xy$ plane in (c). The scale bars denote 20
      $\mu$m; see also \cite{footnoteMovies}.}
      \label{figFingering}
  \end{figure}
\end{center}

In order to drive the system further out-of-equilibrium we have
altered the initial state by turning equilibrated samples upside-down,
such that the colloids preferentially reside at the top of the
capillary. For this system $\xig\approx 0.35\sigma$ and
$L\approx18\sigma$, such that ${\rm Pe}\approx1.6$. Despite the fact
that these parameters are comparable to those considered above,
sedimentation proceeds in an entirely different fashion. A strong
degree of lateral inhomogeneity develops \cite{simulationsPrivCom},
see Fig.~\ref{figFingering}. In the horizontal ($xy$) plane, see Fig.\
\ref{figFingering}c, a network-like structure emerges
\cite{footnoteMovies}. The structure formation is reminiscent of that
resulting from the Rayleigh-Taylor instabililty in two layered
liquids, given that the steep initial density gradient resembles that
of a liquid-liquid interface. There is likely a relationship between
this phenomenon and the swirls observed in sedimenting granular matter
\cite{segre2001}.
A theoretical investigation of this instability requires an explicit
treatment of the lateral $x$ and $y$ coordinates and a more
sophisticated handling of the solvent hydrodynamics \cite{capuani2004}
than only through a density-dependent mobility; these tasks are beyond
the scope of the present work.  We believe that DDFT, when extended
along these lines, could be a very promising candidate for studying
such phenomena, both on the grounds of the high accuracy in simpler
situations (demonstrated above) and that it also has been used
successfully in the past to describe pattern formation in processes
such as spinodal decomposition \cite{archer2005}.


In conclusion we have presented the first measurements on the
single-particle level of sedimentation in model colloidal
dispersions. Unlike previous macroscopic descriptions that apply when
the gravitational length is much larger than the particle size
\cite{janDhont}, we have considered cases where both lengths are
similar in magnitude. We have demonstrated that DDFT gives accurate
results as compared to computer simulation results and that it
describes laterally homogenous sedimentation very well as compared to
experimental data. Besides systems that exhibit lateral pattern
formation it would be interesting to investigate in future work larger
systems where crystallization occurs as well \cite{dullens06}.

\begin{acknowledgments}
We thank A.J.\ Archer and H.\ Tanaka for many illuminating
discussions, H.\ Tanaka for microscope time, and D.\ Derks for
particle synthesis. J.D.\ and M.S.\ thank the \emph{Deutsche
Forschungsgemeinschaft} for support via the Emmy-Noether-Programme and
SFB-TR6/D3, respectively. C.P.R. thanks the {\em Japan Society for the
Promotion of Science} for support.  This work is part of the
\emph{Stichting voor Fundamenteel Onderzoek der Materie} (FOM), which
is supported by the \emph{Nederlandse Organisatie voor
Wetenschappelijk Onderzoek} (NWO).
\end{acknowledgments}

\clearpage

\begin{table}
  \begin{tabular}{|l|ccccccccc|}\hline
    & 
    $\phi_0$ & $\nu\sigma^2$ & $\sigma$ & $L$ & 
    $\xig/\sigma$ & $L/\sigma$ & $\tau_B$ (s) & $\sigmaeff/\sigma$ & Pe \\
    \hline
    A & 
    0.15 & 3.624 & 3.5 & 50 & 0.8 & 12(+1) & 57.9 & 1.05 & 0.625\\
    B & 
    0.142 & 10.46 & 2.8 & 100 & 0.45 & 38.5(+1) & 29.3 & 1.1 & 1.11\\
    \hline
  \end{tabular} 
  \caption{Overview of parameters for systems A and B: Total hard
    sphere packing fraction $\phi_0$, scaled density per lateral area
    $\nu\sigma^2$, where $\phi_0 = \pi\sigma^3\nu/(6L)$, particle
    diameter $\sigma$ (in microns) as obtained from static light
    scattering, capillary spacing $L$ (in microns), scaled
    gravitational length $\xig/\sigma$, scaled plate separation
    distance $L/\sigma$, Brownian time $\tau_B$ (in seconds), scaled
    effective (electrostatic) hard sphere diameter $\sigmaeff/\sigma$,
    and Peclet number Pe.  The values for $\xig/\sigma$ and
    $\sigmaeff/\sigma$ are obtained by fitting theoretical results for
    equilibrium density profiles of hard spheres to those from
    experiment. The lateral size of the capillaries was 1 mm $\times$
    50 mm.  }
  \label{table1}
\end{table}

\clearpage

\clearpage

\end{document}